# A Pc-Kriging-Hdmr Integrated with an Adaptive Sequential Sampling Strategy for High-Dimensional Approximate Modeling

Yili Zhang, Hanyan Huang*, Mei Xiong, Zengquan Yao

School of Systems Science and Engineering, Sun Yat-Sen University,
Guangzhou, Guangdong 510275, PR China

## Abstract

*High-dimensional complex multi-parameter problems are ubiquitous in engineering, as traditional surrogate models are limited to low/medium-dimensional problems (referred to $p \leq 10$ ). They are limited to the dimensional disaster that greatly reduce the modelling accuracy with the increase of design parameter space. Moreover, for the case of high nonlinearity, the coupling between design variables fail to be identified due to the lack of parameter decoupling mechanism. In order to improve the prediction accuracy of high-dimensional modeling and reduce the sample demand, this paper considered embedding the PC-Kriging surrogate model into the high-dimensional model representation framework of Cut- HDMR. Accordingly, a PC-Kriging-HDMR approximate modeling method based on the multi-stage adaptive sequential sampling strategy (including first-stage adaptive proportional sampling criterion and second stage central-based maximum entropy criterion) is proposed. This method makes full use of the high precision of PC-Kriging and optimizes the layout of test points to improve the modeling efficiency. Numerical tests and a cantilever beam practical application are showed that: (1) The performance of the traditional single-surrogate model represented by Kriging in high-dimensional nonlinear problems is obviously much worser than that of the combined- surrogate model under the framework of Cut-HMDR (mainly including Kriging-HDMR, PCE-HDMR, SVR-HDMR, MLS-HDMR and PC-Kriging-HDMR);(2)The number of samples for PC-Kriging-HDMR modeling increases polynomially rather than exponentially with the expansion of the parameter space, which greatly reduces the computational cost;(3)None of the existing Cut-HDMR can be superior to any in all aspects. Compared with PCEHDMR and Kriging-HDMR, PC-Kriging-HDMR has improved the modeling accuracy and efficiency*
*within the expected improvement range, and also has strong robustness.*

## 1. Introduction

Approximate modelling is an integrated application of experimental design, mathematical statistics and optimization techniques. It's principle is to construct a simplified mathematical model which meets the accuracy requirements to simulate the input-output relationship of the problem to replace the complex and time-consuming original model. It can realize the purpose of accelerating design and improving analysis efficiency [1]. It's precisely because of the excellent characteristics of approximate modelling such as repeatability and foresight that it has aroused strong research interest. [2]. However, most of the existing approximate models, such as Radial Basis Function (RBF), Kriging and Polynomial Chaotic Expansion (PCE), are limited to solve high-dimensional problems (referred to $p>10$) [3]. Specifically, there are two main bottlenecks: (1) With the rapid expansion of the design space, the computational cost increases exponentially, which makes it difficult to obtain expected high-precision approximate models; (2) The lack of parameter decoupling mechanism makes it hard to reveal the essence of capturing the coupling relationship between variables. As a representative method to solve high-dimensional problems, High Dimensional Model Representation (HDMR) [4] can clarify the coupling characteristics between input variables and construct a global surrogate model, which has a very broad application prospect.

HDMR is a multi-parameter decoupling technique, which was firstly proposed by Sobol in 1993 [4]. Rabitz [5] and Li [6] subsequently perfected the theory: For any integrable function defined in the domain, there is a unique high-dimensional model representation (i.e., HDMR multi-parameter decoupling expression), which lays the foundation for the establishment of surrogate model under high-dimensional problems. There are two derivative forms of HDMR: ANOVA-HDMR and Cut-HDMR [7]. In particular, Cut-HDMR has been widely used to predict valuable high-dimensional black-box problems due to its high computational efficiency [8]. There are many derivative combinations under the framework of Cut-HDMR, such as Kriging-HDMR, MLS-HDM, RBF-HDMR, SVR-HDMR and PCE-HDMR. Wang [9] proposed a high-dimensional model representation method based on Moving Least Square (MLS-HDMR) in 2011, which combines the DIRECT sampling strategy to construct the meta-surrogate model. Shortly after that Tang [10] proposed the Kriging-HDMR, which designed online adaptive sequential sampling mechanism to iteratively update the component terms of Cut-HDMR. Li [11] proposed SVM-HDMR in 2013 by embedding Least Squares Support Vector Machine (LS-SVM) into the framework of Cut-HDMR. In 2015, Qiu's team [12] proposed the AERBF-HDMR (Adaptive ESO-based RBF-HDMR) by combining the LORA-Voronoi sequential sampling strategy to obtain an accurate global surrogate model with less number of model evaluations. In 2016, based on the Bisection sampling method, Yu [13] proposed an improved SVR-HDMR model, called BS-SVR-HDMR model, which effectively deal with the sparsity and complexity in high-dimensional nonlinear models. Subsequently, Cai et al. [14] proposed MFHDMR (Multi-Fidelity based Cut-HDMR), which combined multi-fidelity modeling (Kriging and Co-Kriging) with Cut-HDMR. It performs better in the case of diverse data and missing data. Literature [15] comprehensively compared the accuracy and efficiency of HDMRs in high-dimensional modeling, and concluded that the performance of Kriging-HDMR was slightly better than RBF-HDMR and SVR-HDMR. The above methods can identify variable coupling relationships, and then reduce model complexity and sample requirements when dealing with high-dimensional problems [16]. However, the performance of above methods mostly depends on the performance of single-surrogate models such as PCE, Kriging, RBF, and the accuracy and sampling strategies need to be further optimized.

This paper considered embedding the PC-Kriging surrogate model into the high-dimensional model representation modeling framework of Cut-HDMR to propose a new approximate modeling method, named PC-Kriging-HDMR. Accordingly, a multi-stage adaptive sequential sampling strategy is combined with it to reduce sample demand. The remaining sections are arranged as follows: **Section 2** introduces the basic principle of PC-Kriging-HDMR; In **Section 3**, the sequential sampling strategy is elaborated, and the specific construction process of the decoupled model (PC-Kriging-HDMR) is given. In **Section 4**, a variety of numerical experiments are carried out to compare the performance of modling accuracy and efficiency with a series of Cut-HDMRs. In **Section 5**, the "Stress-strength Interference Model" is used to modeling the reliability of the cantilever, and PC-Kriging-HDMR is used for sensitivity analysis. Concluding remarks are summarized in **Section 6**.

## 2. THE BASIC PRINCIPLE OF PC-KRIGING-HDMR

### 2.1. Cut-based High Dimensional Model Representation (Cut-HDMR)

High Dimensional Model Representation (HDMR) [4] decomposes a multi-parameter function into the sum of several functions with fewer parameters according to the coupling between variables. Set $\Omega^p = \{(x_1, x_2, \ldots, x_p) | 0 \leq x_i \leq 1, i = 1, 2, \ldots, p\}$ be a subset of the $p$ dimensional Euclidean space $R^p$, and $\mathrm{X} = \{f | f(X) = f(x_1, x_2, \ldots, x_p), X \in \Omega^p\}$ is the linear space of all functions defined on $\Omega^p$. Define a measure $\mu$ on $\Omega^p$ such that:

$$\begin{cases} d\mu = d\mu(x_1, x_2, \ldots, x_p) = \prod_{i=1}^{p} d\mu_i(x_i) \\ \int_{\Omega^i} d\mu_i(x_i) = 1 (i = 1, 2, \ldots, p) \\ d\mu(X) = \mathbf{g}(X)dX = \prod_{i=1}^{p} g_i(x_i) dx_i \end{cases} \quad (1)$$

The inner product on the linear space X can be defined by the measure $\mu$ as follows:

$$\langle f, h \rangle = \int_{\Omega_p} f(X) h(X) d\mu(X), \quad f(X), h(X) \in \mathrm{X} \quad (2)$$

It can be shown from [4,7] that X can be the direct sum of the specific subspaces, i.e.,

$$\mathrm{X} = X_0 \oplus \sum_{1 \leq i \leq p} X_i \oplus \sum_{1 \leq i < j \leq p} X_{ij} \oplus \cdots \sum_{1 \leq i_1 < i_2 < \ldots, i_j \leq p} X_{i_1, i_2 \ldots, i_j} \oplus X_{1,2,\cdots,p} \quad (3)$$

Based on the above theoretical, HDMR is described as follows: Suppose the output response corresponding to the input variable $X = [x_1, x_2, \ldots, x_p]^T$ in the $p$ dimensional design space is $f(X)$, and the mapping relationship between them is expressed as:

$$\begin{aligned} f(X) &= f_0 + \sum_{i=1}^{p} f_i(x_i) + \sum_{1 \leq i < j \leq p} f_{ij}(x_i, x_j) + \sum_{1 \leq i < j < k \leq p} f_{ijk}(x_i, x_j, x_k) + \cdots \\ &+ \sum_{1 \leq i_1 < \cdots < i_l \leq p} f_{i_1 i_2 \cdots i_l}(x_{i_1}, x_{i_2}, \cdots, x_{i_l}) + \cdots + f_{1,2,\cdots,p}(x_1, x_2, \cdots, x_p) \end{aligned} \quad (4)$$

Cut-HDMR is to pre-determine the cut-center point $\mathbf{x}_0 = (c_1, c_2, c_3, \ldots, c_p)$, and the high-dimensional modal expression of $f(X)$ is formalised as the sum of a series of lines, planes and hyperplanes that pass through the cut-center point. The expressions for the tuple terms of each order of Cut-HDMR are as follows:

$$\begin{aligned} f_0 &= f(\mathbf{x}_0) \\ f_i(x_i) &= f(x_i, x_0^i) - f_0 \\ f_{ij}(x_i, x_j) &= f(x_i, x_j, x_0^{ij}) - f_i(x_i) - f_j(x_j) - f_0 \\ &\ldots \\ f_{12\ldots p}(x_1, x_2, \ldots, x_p) &= f(\mathbf{x}) - \sum_i f_i(x_i) - \sum_{ij} f_{ij}(x_i, x_j) - \ldots - f_0 \end{aligned} \quad (5)$$

Where, $f_0$ is the output response of the original model at $\mathbf{x}_0 = (c_1, c_2, c_3, \ldots, c_p)$. $(x_i, x_0^i)$ is $(c_1, c_2, \ldots, c_{i-1}, x_i, c_{i+1}, \ldots, c_p)$, whose entries are equal to the values corresponding to the $\mathbf{x}_0$, except for the independent variable $x_i$. In this vein, the central basis of higher order terms has the similar meaning, as shown in **Figure 1**:

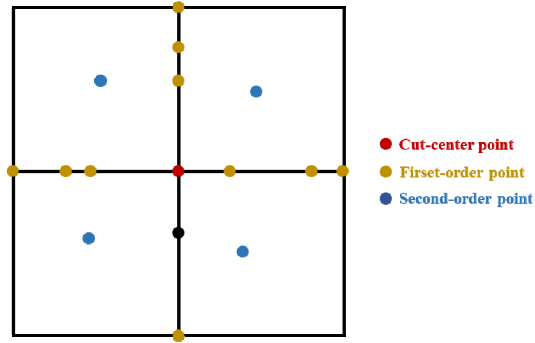

**Figure 1.** Cut-HDMR sampling distribution diagram

Reference [5] pointed out that the expansion form of Cut-HDMR is the optimal decomposition of $f(X)$ in the $p$ dimensional space, so the total approximation error $\|f(X)-f_{\text{model}}(X)\|^2$ is also minimum.

## 2.2 PC-Kriging

The Polynomial-Chao-based Kriging (PC-Kriging) [17-18] is a new surrogate model that making full use of the global prediction performance of PCE [19] and the local fitting effect of Kriging [20]. PC-Kriging is to expand the probability distribution function of the input random variables into a linear combination of a set of orthogonal polynomials, and embed the optimal set of polynomials into the Kriging model as its global approximation function, while the stochastic process of Kriging captures the randomness in the residual term. PC-Kriging is an efficient, accurate and reliable surrogate model. It has a wide range of applications in dealing with the uncertainty of input random variables and providing uncertainty analysis.

Consider a black-box model $y = M(X)$, where $X$ is a $p$-dimensional with independent variables, $\Omega = \Omega_1 \times \Omega_2 \times \cdots \times \Omega_p$ is the input space, $f_X(x_1, x_2, \cdots, x_p)$ is the joint probability density function of $X$. PCE approximates the black-box model $y = M(X)$ as follows:

$$y = M(X) \approx M^{\text{PCE}}(X) = \sum_{\alpha \in \mathcal{A}} y_\alpha \Phi_\alpha(X) \tag{6}$$

where $\Phi_\alpha(X)$ is a multivariate orthogonal polynomial, $\mathcal{A} \subset \mathbf{N}^p$ is the set of $\alpha$, and $y_\alpha$ is the coefficient corresponding to each multivariate orthogonal polynomial. Kriging approximates the black-box model $y = M(X)$ as follows:

$$y = M(X) \approx M^{(K)}(X) = \boldsymbol{\beta}^T f(X) + Z(X) \tag{7}$$

where $\boldsymbol{\beta}^T f(X)$ is a global approximation function (trend function), coefficient $\boldsymbol{\beta}$ is a parameter to be estimated, and $Z(X)$ represents a stationary gaussian process with zero mean. Combining PCE and Kriging, PC-Kriging is described:

$$y = M(X) \approx M^{\text{PC-Kriging}}(X) = \sum_{\alpha' \in \mathcal{A}'} y_{\alpha'} \Phi_{\alpha'}(X) + Z(X) \tag{8}$$

Specifically, PC-Kriging can be combined in two ways: Sequential PC-Kriging and Optimal PC-Kriging. The main difference between the two is that SPC-Kriging directly uses LARs algorithm to obtain the optimal polynomial set as the orthogonal polynomial basis function of

International Journal of Computer Science & Information Technology (IJCSIT) Vol 15, No 3, June 2023

PCE, and then substitutes it into the trend term of Kriging; The OPC-Kriging is to gradually establish the Kriging with the PCE trend term in the above optimal polynomial set, and take the polynomial set that minimizes the overall model error as the orthogonal polynomial basis function of PCE, then the corresponding PC-Kriging model is the final surrogate model.

## 2.3. PC-Kriging-HMDR

In this section, the PC-Kriging surrogate model is used to construct the Cut-HDMR component subfunctions, which is expressed as follows:

$$\begin{cases} f_i(x_i) \approx M^{\text{PC-Kriging}}(x_i) = \sum_{\alpha' \in \mathcal{A}'} y_{\alpha'} \Phi_{\alpha'}(x_i) + Z(x_i); \\ f_{ij}(x_i, x_j) \approx M^{\text{PC-Kriging}}(x_i, x_j) = \sum_{\alpha' \in \mathcal{A}'} y_{\alpha'} \Phi_{\alpha'}(x_i, x_j) + Z(x_i, x_j); \\ \cdots\cdots \end{cases} \quad (9)$$

Thus, the general expression of PC-Kriging-HDMR is obtained as equation (10):

$$f(\mathbf{x}) \approx f_0 + \overbrace{\sum_{i=1}^{p} \left( \sum_{\alpha' \in \mathcal{A}'} y_{\alpha'} \Phi_{\alpha'}(x_i) + Z(x_i) \right)}^{\text{first-order uncoupled term}} + \overbrace{\sum_{1 \le i < j \le p} \left( \sum_{\alpha' \in \mathcal{A}'} y_{\alpha'} \Phi_{\alpha'}(x_i, x_j) + Z(x_i, x_j) \right)}^{\text{second-order coupled term}} \\ + \cdots + \overbrace{\sum_{\alpha' \in \mathcal{A}'} y_{\alpha'} \Phi_{\alpha'}(x_1, \ldots, x_p) + Z(x_1, \ldots, x_p)}^{p\text{-order coupled term}} \quad (10)$$

In the analysis of most practical engineering problems, the uncoupled terms and the lower order coupling terms are sensitive to the response function. To this end, the constructed PC-Kriging-HDMR is truncated to the second-order coupling terms:

$$f(\mathbf{x}) \approx f_0 + \sum_{i=1}^{p} \hat{f}_i(x_i) + \sum_{1 \le i < j \le p}^{p} \hat{f}_{ij}(x_i, x_j) \quad (11)$$

## 3. PC-KRIGING-HDMR COMBINED WITH MULTI-STAGE ADAPTIVE SEQUENTIAL SAMPLING STRATEGY

### 3.1. Multi-stage adaptive sampling strategy

**(1) First-stage adaptive proportional sampling criterion**

When new sample points are needed to update the PC-Kriging of the first-order uncoupled term, the constructed sample point set $X_i = \{x_{i\_lower}, x_0, x_{i\_upper}\ldots\}$ is arranged in ascending order to obtain $\tilde{X}_i = \{r_{i1}, r_{i2}, \ldots, r_{in}\}$, where $r_{i1} = x_{i\_lower}$ and so on. By using the principle of interval sampling with large nonlinearity to speed up convergence, find two adjacent points $r_{ik}$ and $r_{i(k+1)}$, which measure the degree of nonlinearity is the largest interval:

$$\Delta f = \max_{k=1,2,\ldots,n} \left( \left| f_i(r_{i(k+1)}) - f_i(r_{ik}) \right| \right) \quad (12)$$

Then a new sample point $x_{inew}$ is inserted in this interval:

$$x_{inew} = C r_{ik} + (1-C) r_{i(k+1)} \quad (13)$$

where $C$ is the scale coefficient. If $\left| \dfrac{\hat{f}(x_i) - f(x_i)}{f(x_i)} \right| \le \varepsilon$ (usually regarded as $\varepsilon = 10^{-3}$) is satisfied or the number of existing used samples exceeds a given threshold, the first-order uncoupled term $\hat{f}_i(x_i)$ construction terminates; Otherwise, sampling continues until the convergence criterion is satisfied.

**(2) Second-stage central-based maximum entropy criterion**

Points are firstly selected from the existing candidate set to update the second-order coupling terms. If the candidate set points have been selected, new sample points are constructed by sequential sampling based on central-based maximum entropy criterion. The so-called maximum entropy [21] is defined as the maximum amount of information obtained by an experimental design, that is the determinant of the prior covariance matrix of the sample point set $D_{X_{ij}}$:

$$\max \ \det\left[\operatorname{cov}\left(D_{X_{ij}}, D_{X_{ij}}\right)\right] \tag{14}$$

where $\operatorname{cov}\left(D_{X_{ij}}, D_{X_{ij}}\right) = \sigma^2 \Re\left[R\left(d_{x_{ij}}, d_{x_{ij}}'\right)\right], d_{x_{ij}}, d_{x_{ij}}' \in D_{X_{ij}}$. $\sigma^2$ represents the prior variance of the sample points, and $\Re$ is the correlation matrix formed by the prior covariance function $R$. The prior covariance function chosen in this paper is Gaussian, i.e.:

$$R\left(d_{x_{ij}}, d_{x_{ij}}'\right) = \exp\left[-\sum_{k=1}^{p} \theta_k \left|\left(d_{x_{ij}}\right)^k - \left(d_{x_{ij}}'\right)^k\right|^2\right] \tag{15}$$

The relevant parameter vector $\boldsymbol{\theta}$ can be obtained by optimizing the following formula:

$$\max \ -\frac{1}{2}\left[N \ln\left(\hat{\sigma}^2\right) + \ln|\Re|\right] \tag{16}$$

Therefore, for the existing sample set $D_{X_{ij}(N)} = \left\{d_{x_{ij}(1)}, d_{x_{ij}(2)}, d_{x_{ij}(3)}, ..., d_{x_{ij}(N)}\right\}$, select the new sample point $d_{x_{ij}(N+1)}$ to maximize its entropy:

$$\max \ \det\left[\operatorname{cov}\left(D_{X_{ij}(N+1)}, D_{X_{ij}(N+1)}\right)\right] \tag{17}$$

where $\operatorname{cov}\left(D_{X_{ij}(N+1)}, D_{X_{ij}(N+1)}\right) = \sigma^2 \begin{bmatrix} \operatorname{cov}\left(D_{X_{ij}(N)}, D_{X_{ij}(N)}\right) & r\left(d_{x_{ij}(N+1)}\right) \\ r\left(d_{x_{ij}(N+1)}\right)^T & 1 \end{bmatrix}$ and satisfies:

$$r\left(d_{x_{ij}(N+1)}\right) = \left[R\left(d_{x_{ij}(N+1)}, d_{x_{ij}(1)}\right), R\left(d_{x_{ij}(N+1)}, d_{x_{ij}(2)}\right), R\left(d_{x_{ij}(N+1)}, d_{x_{ij}(3)}\right), ..., R\left(d_{x_{ij}(N+1)}, d_{x_{ij}(N)}\right)\right] \tag{18}$$

Specifically, the plane parameter space is divided into several small matrices by using the sampled data point set in the first stage, and then a point that maximizes the system entropy at the cut-center point of each matrix is selected as a new sample point for constructing the second-order couple term. The above steps are repeated until the convergence criterion is satisfied. This sequential sampling strategy based on maximum entropy criterion can obtain comprehensive uncertainty information in the explored parameter space, and has higher diversity and robustness of modeling.

### 3.2. The modeling process of PC-Kriging-HDMR

The specific process of PC-Kriging-HDMR approximate modeling combined with multi-stage adaptive sequential sampling strategy is described detailly in this section. The proposed multi-stage adaptive sampling strategy utilizes the HDMR hierarchy and divides the sampling domain into multiple subdomains, and properly samples in each subdomain to reduce the sampling cost, thus avoiding the difficulty of high-dimensional modeling. In addition, the sampling strategy also reveals the characteristics of design variables, including linearity and nonlinearity, coupling, etc., which directly affect the predictive ability and stability of the model.

The flow chart of the algorithm is shown in **Figure 2**.

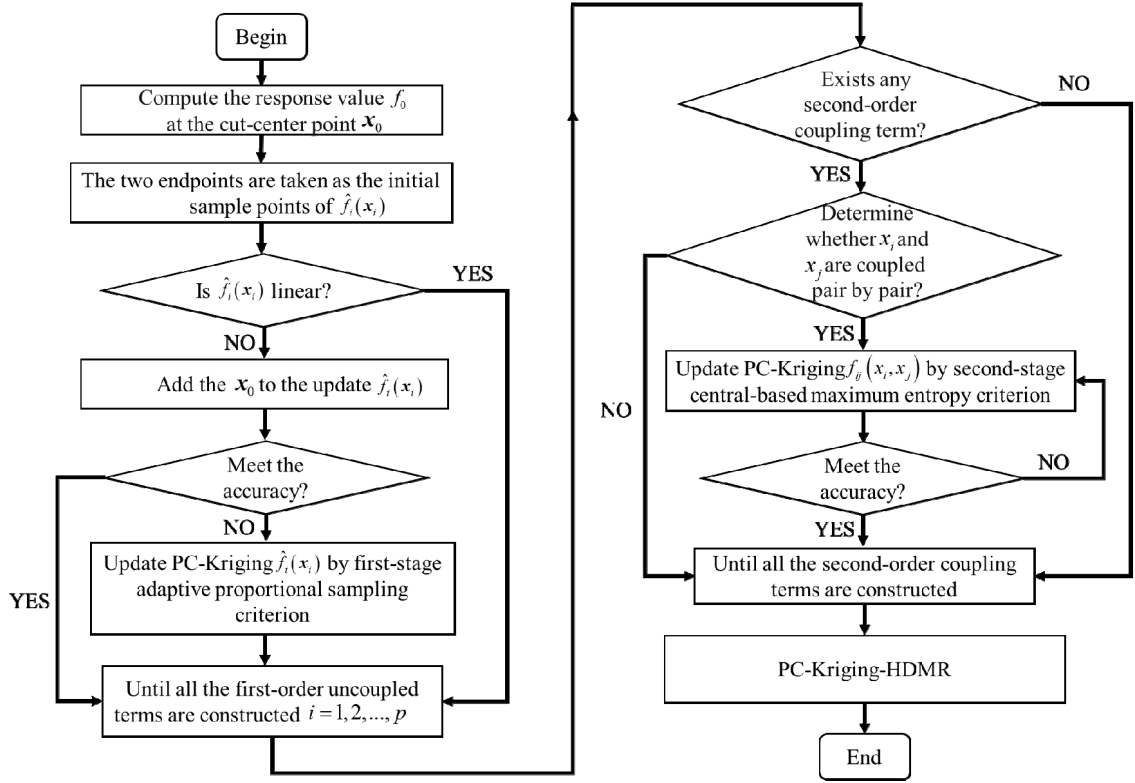

**Figure 2.** Flow chart of PC-Kriging-HDMR combined with multi-stage adaptive sequential sampling strategy

(1) Take an arbitrary point $x_0 = (c_1, c_2, c_3, ..., c_p)$ as the cut-center point. Then compute the response of the true output function at $x_0$ to get $f_0$.

(2) Construct PC-Kriging surrogate models consisting of first-order uncoupled terms dimensionally. The points of $f_i(x_i) = f\left(\left[c_1, c_2, ..., c_{i-1}, x_i, c_{i+1}..., c_p\right]^T\right) - f_0$ are arranged on the value interval of the single variable $x_i\left(\left[c_1, c_2, ..., c_{i-1}, x_i, c_{i+1}..., c_p\right]^T\right)$. That is, two sample points are randomly generated in the upper and lower boundary neighborhood of the $i$-th dimension design variable, and their target response values are calculated:

$$f(x_{i\_lower}) = f\left(\left[c_1, c_2, ..., c_{i-1}, x_{i\_lower}, c_{i+1}..., c_p\right]^T\right) - f_0;$$

$$f(x_{i\_upper}) = f\left(\left[c_1, c_2, ..., c_{i-1}, x_{i\_upper}, c_{i+1}..., c_p\right]^T\right) - f_0.$$

Then $\hat{f}_i(x_i)$ is constructed by constructing the corresponding PC-Kriging surrogate model using the above two sampling points.

**(3)** Determine whether $f_i(x_i)$ is linear or non-linear. If the initial Kriging-surrogate model $\hat{f}_i(x_i)$ passes through the cut-center point $x_0$, it is regarded as a linear term and the construction terminates. Otherwise, $f_i(x_i)$ is a nonlinear term and proceed to the next step.

**(4)** Construct nonlinear first-order terms based on adaptive proportional sampling criterion in the first stage.

**(5)** Iterate through steps (2) to (4) until all first-order uncoupled terms (for each $i=1,2,...,p$) are constructed.

**(6)** Determine whether there is any second-order coupled term. Construct a new sample point $\boldsymbol{x}_e = \left[x_{e_1}, x_{e_2}, ..., x_{e_i}, ..., x_{e_j}, ..., x_{e_p}\right]^T$ and without loss of generality, one of the sample point components $x_{i\_lower}$ and $(i=1,...,p)$ used in the construction of the first-order uncoupled term is randomly chosen as the $i$-th dimensional component $x_{e_i}$ of the new sample point. Within the error range allowed by the accuracy criterion, if $f(\boldsymbol{x}_e) = f_0 + \sum_{i=1}^{p}\hat{f}_i(x_{e_i})$ is satisfied, it is considered that there is no second-order coupled term in the model, and the construction process is finished. Otherwise, proceed to step (7).

**(7)** Construct the second-order coupled terms based on central-based maximum entropy criterion in the second stage, until all second-order variable combinations have been identified.

**(8)** The high-order coupled functions contribute little so that PC-Kriging-HDMR is constructed to the second-order coupled terms.

## 4. NUMERICAL SIMULATION EXPERIMENT

### 4.1. Experimental evaluation index

The performance indexes $R^2$, RAAE and RMAE mainly consider the approximate accuracy and modeling efficiency of the models.

**(1)** The judgment coefficient, $R^2$

$$R^2 = 1 - \frac{\sum_{i=1}^{s}(f(X_i) - \hat{f}(X_i))^2}{\sum_{i=1}^{s}(f(X_i) - \bar{f})^2} \tag{18}$$

**(2)** Relative Average Absolute Error, RAAE

$$RAAE = \frac{\sum_{s}^{i=1}|f(X_i) - \hat{f}(X_i)|}{s \cdot STD} \tag{19}$$

where, $STD = \sqrt{\frac{1}{s-1}\sum_{s}^{i=1}(f(X_i) - \bar{f})^2}$ is the standard deviation.

**(3)** Relative Maximum Absolute Error, RMAE

$$RMAE = \frac{\max\left\{|f(f(X_i)) - \hat{f}(f(X_i))|,...,|f(f(X_s)) - \hat{f}(f(X_s))|\right\}}{STD} \tag{20}$$

## 4.2. Proportionality coefficient test

As one of the strengths of PC-Kriging-HDMR is identifying coupling relationships between variables, the coupling test function is selected the non-convex Rosenbrock function proposed by Howard Harry Rosenbrock in 1960:

$$f(\boldsymbol{x}) = \sum_{i=1}^{8} \left[ 100\left(x_{i+1} - x_i^2\right)^2 + \left(x_i - 1\right)^2 \right], -2 \leq x \leq 2 \qquad (21)$$

The scale coefficient $C = \{0.1, 0.2, 0.3, 0.4, 0.5, 0.6, 0.7, 0.8, 0.9\}$ were tested for 10 times to compare. The coupling test results are shown in **Table 1**, and modelling accuracy results are shown in **Figure 3** and **Table 2**:

Table 1. Coupling test results

|       | $x_1$ | $x_2$ | $x_3$ | $x_4$ | $x_5$ | $x_6$ | $x_7$ | $x_8$ | $x_9$ |
|-------|---|---|---|---|---|---|---|---|---|
| $x_1$ | 1 | 1 | 0 | 0 | 0 | 0 | 0 | 0 | 0 |
| $x_2$ | 1 | 1 | 1 | 0 | 0 | 0 | 0 | 0 | 0 |
| $x_3$ | 0 | 1 | 1 | 1 | 0 | 0 | 0 | 0 | 0 |
| $x_4$ | 0 | 0 | 1 | 1 | 1 | 0 | 0 | 0 | 0 |
| $x_5$ | 0 | 0 | 0 | 1 | 1 | 1 | 0 | 0 | 0 |
| $x_6$ | 0 | 0 | 0 | 0 | 1 | 1 | 1 | 0 | 0 |
| $x_7$ | 0 | 0 | 0 | 0 | 0 | 1 | 1 | 1 | 0 |
| $x_8$ | 0 | 0 | 0 | 0 | 0 | 0 | 1 | 1 | 1 |
| $x_9$ | 0 | 0 | 0 | 0 | 0 | 0 | 0 | 1 | 1 |

Table 2. The median results corresponding to Figure 3

| C   | $R^2$ | RAAE  | RMAE  |
|-----|-------|-------|-------|
| 0.1 | 0.672 | 0.433 | 2.424 |
| 0.2 | 0.673 | 0.429 | 2.416 |
| 0.3 | 0.676 | 0.428 | 2.411 |
| 0.4 | 0.681 | 0.423 | 2.332 |
| **0.5** | **0.692** | **0.414** | **2.113** |
| 0.6 | 0.679 | 0.419 | 2.314 |
| 0.7 | 0.678 | 0.422 | 2.467 |
| 0.8 | 0.675 | 0.425 | 2.532 |
| 0.9 | 0.870 | 0.426 | 2.748 |

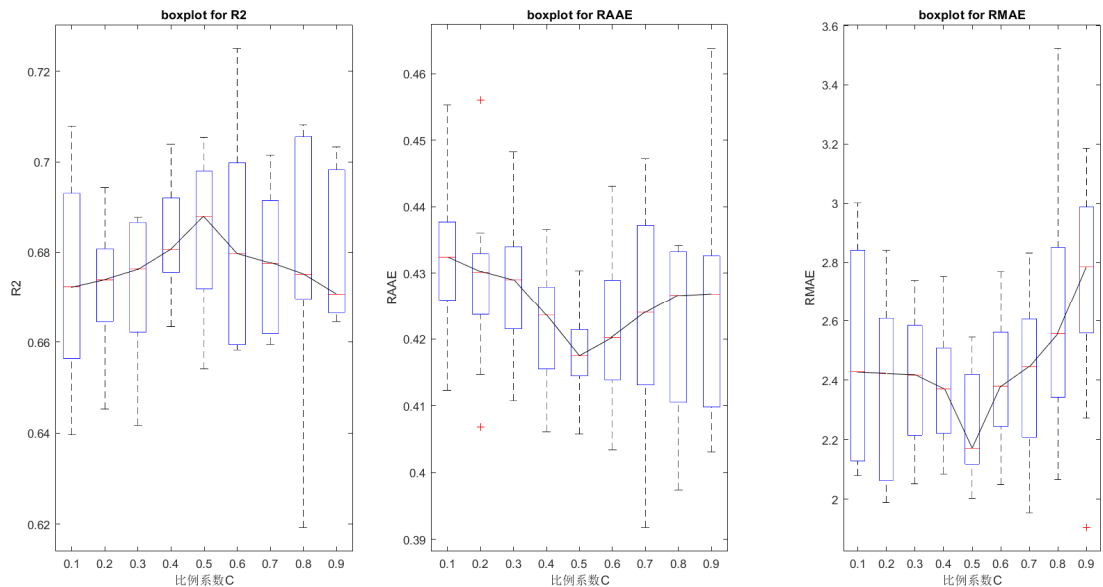

Figure 3. Box graph of accuracy results of 10 tests for Rosenbrock function

Table 1 shows that PC-Kriging-HDMR passes the coupling test, which correctly identify the variable coupling relationship of the Rosenbrock test function. According to the Table 2 and Figure 3, it can be seen that when the proportion coefficient C=1/2, the performance is the best, and the closer the proportion coefficient is to the middle value 0.5, the better the approximate ability of the positive distribution trend. That is, the maximum $R^2$ is 0.692, the RAAE and RMAE are 0.414 and 2.113 respectively, which are the smallest. Therefore, under the same computational cost, PC-Kriging-HDMR has higher global approximation accuracy and stronger robustness when the scaling coefficient C is 1/2.

### 4.3. High-dimensional nonlinear test

This test is mainly focuses on high-dimensional nonlinear functions, 6 test functions (including 3 low/medium-dimensional functions and 3 high-dimensional functions) are cited in [12] to compare with single-surrogate model Kriging and five combined-surrogate models including Kriging-HDMR, PCE-HDMR, SVR-HDMR, MLS-HDMR and PC-Kriging-HDMR. The UQLab toolbox is used to build the surrogate model, and the test functions are shown in **Table 3**. The test results are shown in **Table 4**.

The results show that the performance of traditional single-surrogate models represented by Kriging is obviously much worser than that of Cut-HDMRs in high-dimensional nonlinear problems. In particular, the $R^2$ of combined-HDMRs series methods remain above 0.95, while the performance of Kriging is extremely unstable, even as low as 0.2. For the series of multi-parameter decoupling models based on surrogate model (mainly including Kriging-HDMR, PCE-HDMR, SVR-HDMR, MLS-HDMR and PC-Kriging-HDMR), PC-Kriging-HDMR has the best performance on test functions NO.1,2,5,6 by absolute advantage. It shows that the PC-Kriging improved by combining Kriging and PCE is within the expected improvement range in $R^2$, RAAE and RMAE using adaptive sequential strategy in the multi-parameter decoupling modeling framework, which has wider applicability and stronger approximation ability for high-dimensional complex problems

**Table 3.** High-dimensional nonlinear test functions

| NO. | Test functions | Design space |
|---|---|---|
| 1 | $f(\boldsymbol{x}) = \sin(x_1 + x_2) + (x_1 - x_2)^2 - 1.5x_1 + 2.5x_2 + 1$ | $x_i \in [-3,3], i = 1,2$ |
| 2 | $f(\boldsymbol{x}) = \left(x_2 - 1.275\left(\frac{x_1}{\pi}\right)^2 - 5\frac{x_1}{\pi} - 6\right)^2 + 10\left(1 - \frac{0.125}{\pi}\right)\cos(x_1)$ | $x_1 \in [-5,10], x_2 \in [0,15]$ |
| 3 | $f(\boldsymbol{x}) = \sum_{i=1}^{8-1}\left((x_{i+1}^2 - x_i)^2 + (x_i - 1)^2\right)$ | $x_i \in [-3,3], i = 1,...,8$ |
| 4 | $f(\boldsymbol{x}) = x_1^2 + x_2^2 + x_1x_2 - 14x_1 - 16x_2 + (x_3 - 10)^2 + 4(x_4 - 5)^2 + (x_5 - 3)^2$ $+ 2(x_6 - 1)^2 + 5x_7^2 + 7(x_8 - 11)^2 + 2(x_9 - 10)^2 + (x_{10} - 7)^2 + 45$ | $x_i \in [-10,11], i = 1,...,10$ |
| 5 | $f(\boldsymbol{x}) = \sum_{i=1}^{10} x_i \left(c_i + \ln\left(\frac{x_i}{\sum_{k=1}^{10}(x_k)}\right)\right),$ $\left(\begin{array}{l}c_{i=1,...,10} = -6.089, -17.164, -34.054, -5.914, -24.721,\\ -14.986, -24.100, -10.708, -26.662, -22.179.\end{array}\right)$ | $x_i \in [2.1, 9.9], i = 1,...,10$ |
| 6 | $f(\boldsymbol{x}) = (x_1 - 1)^2 + \sum_{16}^{i=2} i(2x_i^2 - x_{i-1})^2$ | $x_i \in [-5,5], i = 1,...,16$ |

**Table 4.** High-dimensional nonlinear test results

| Test functions | Methods | $R^2$ | RAAE | RMAE |
| --- | --- | --- | --- | --- |
| N0.1 (p=2) | Kriging | 0.9876 | 0.0979 | 0.2475 |
| | SVR-HDMR | 0.9823 | 0.1045 | 0.3588 |
| | MLS-HDMR | 0.9875 | 0.0893 | 0.2444 |
| | PCE-HDMR | 0.9831 | 0.1118 | 0.3210 |
| | Kriging-HDMR | 0.9750 | 0.1245 | 0.4201 |
| | **PC-Kriging-HDMR** | **0.9899** | **0.0885** | **0.2095** |
| N0.2 (p=2) | Kriging | 0.6128 | 0.3691 | 2.1493 |
| | SVR-HDMR | 0.9379 | 0.1893 | 0.8229 |
| | MLS-HDMR | 0.9717 | 0.1350 | 0.4516 |
| | PCE-HDMR | 0.9422 | 0.1958 | 0.5361 |
| | Kriging-HDMR | 0.9424 | 0.1505 | 0.5345 |
| | **PC-Kriging-HDMR** | **0.9912** | **0.0589** | **0.2249** |
| N0.3 (p=8) | Kriging | 0.2440 | 0.6796 | 5.9153 |
| | SVR-HDMR | 0.7966 | **0.3458** | 2.3386 |
| | MLS-HDMR | **0.7979** | 0.3489 | 2.3783 |
| | PCE-HDMR | 0.5162 | 0.6146 | 1.9761 |
| | Kriging-HDMR | 0.7733 | 0.3510 | 2.1088 |
| | **PC-Kriging-HDMR** | 0.7805 | 0.3508 | **1.8695** |
| N0.4 (p=10) | Kriging | 0.2379 | 0.6561 | 4.1065 |
| | SVR-HDMR | 0.9617 | 0.1725 | 0.4952 |
| | MLS-HDMR | **0.9987** | **0.0269** | 0.1071 |
| | PCE-HDMR | 0.9986 | 0.0276 | 0.1047 |
| | Kriging-HDMR | 0.9986 | 0.0279 | 0.1045 |
| | **PC-Kriging-HDMR** | **0.9987** | 0.0271 | **0.1013** |
| N0.5 (p=10) | Kriging | 0.6124 | 0.4679 | 3.0520 |
| | SVR-HDMR | 0.9887 | 0.0856 | 0.4095 |
| | MLS-HDMR | 0.9974 | 0.0382 | 0.3831 |
| | PCE-HDMR | **1.0000** | 0.0022 | 0.0175 |
| | Kriging-HDMR | **1.0000** | 0.0024 | 0.0219 |
| | **PC-Kriging-HDMR** | **1.0000** | **0.0020** | **0.0172** |
| N0.6 (p=16) | Kriging | 0.2162 | 2.4539 | 5.7049 |
| | SVR-HDMR | 0.9486 | 0.1766 | 1.1416 |
| | MLS-HDMR | 0.9454 | 0.1858 | 1.0978 |
| | PCE-HDMR | 0.9755 | 0.1366 | 0.1002 |
| | Kriging-HDMR | 0.9635 | 0.1457 | 1.0097 |
| | **PC-Kriging-HDMR** | **0.9865** | **0.1044** | **0.9780** |

## 4.4. Computational cost test

The test function $f(x) = \sum_{i=1}^{p-1}\left[\left(x_i^2\right)^{\left(x_{i+1}^2+1\right)} + \left(x_{i+1}^2\right)^{\left(x_i^2+1\right)}\right], 0 \leq x_i \leq 1$ is used to prove the optimized performance of the improved PC-Kriging-HDMR on the calculation cost. $p = 10, 15, ..., 35$ and 7 sampling points were taken for each subitem (meeting $\left|\left(\hat{f} - f\right)/f\right| \leq \varepsilon$). **Table 5** lists the comparison of calculation costs of different levels.

**Table 5.** Test results for calculating cost

| Dimension | PC-Kriging-HDMR | Kriging-HDMR | Full second-order expansion of HDMR $1 + p(s-1) + \frac{p(p-1)}{2}(s-1)^2$ ((polynomial)) |
|---|---|---|---|
| 10 | 125 | 129 | 2276 |
| 15 | 215 | 224 | 5251 |
| 20 | 346 | 355 | 9451 |
| 25 | 482 | 495 | 14876 |
| 30 | 633 | 649 | 21526 |
| 35 | 831 | 868 | 29401 |

From the experimental results, it's proved that the number of samples for PC-Kriging-HDMR modeling increases polynomially rather than exponentially with the expansion of the parameter space, which greatly reduces the computational cost.

## 5. PRACTICAL APPLICATION: A CANTILEVER BEAM

Considering the stress distribution and displacement change of the cantilever under impact load, the "Stress-strength Interference Model" is often used to model the reliability of the cantilever. It's based on two basic assumptions: (1) The strength of the material has a certain probability distribution, which is usually Weibull distribution or Normal distribution; (2) When the structure is under load, its stress also follows a certain probability distribution.

PC-Kriging-HDMR is used to model the data of the "Stress-strength Interference Model", which can evaluate the reliability level of the structure during the design and optimize the structure to improve its reliability. In addition, it helps to understand the influence of different parameters on the reliability of the structure, which can guide the optimization design and decision making. **Figure 4** shows the cantilever beam is subjected to external forces $F_1$、$F_2$、P and torsion T, respectively.

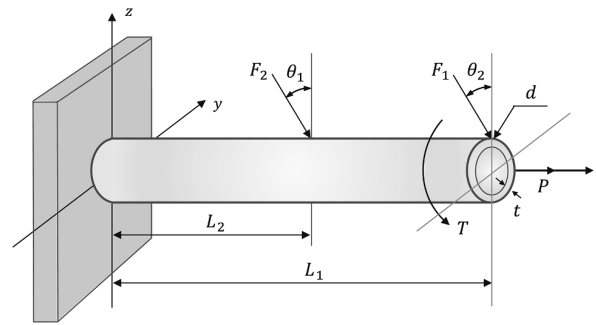

**Figure 4.** Schematic diagram of the cantilever beam model structure

International Journal of Computer Science & Information Technology (IJCSIT) Vol 15, No 3, June 2023

The limit state function is defined in reference [22-23]:

$$G(\boldsymbol{x}) = S_y - \sqrt{\sigma_x^2 + 3\tau_{zx}^2} \quad (22)$$

where $S_y$ is the bending force, $\sigma_x$ and $\tau_{zx}$ are the normal stress and torsional stress at the origin on the top of the tube surface respectively, and their specific expressions are as follows:

$$\begin{cases} \sigma_x = \dfrac{P + F_1 \sin(\theta_1) + F_2 \sin(\theta_2)}{A} + \dfrac{Md}{2I} \\ \tau_{zx} = \dfrac{Td}{2J} \end{cases} \quad (23)$$

The nested subfunction expression in Equation (23) is:

$$\begin{cases} M = F_1 L_1 \cos(\theta_1) + F_2 L_2 \cos(\theta_2) \\ A = (\pi/4)\left[d^2 - (d-2t)^2\right] \\ I = (\pi/64)\left[d^4 - (d-2t)^4\right] \\ J = 2I \end{cases} \quad (24)$$

After hierarchical nesting, the limit state function is obtained from Equations (22), (23) and (24):

$$G(\boldsymbol{x}) = G(t, d, L_1, L_2, F_1, F_2, P, T, \theta_1, \theta_2, S_y)$$

$$= S_y - \left\{\left(\dfrac{P + F_1 \sin(\theta_1) + F_2 \sin(\theta_2)}{(\pi/4)\left[d^2 - (d-2t)^2\right]} + \dfrac{\left[F_1 L_1 \cos(\theta_1) + F_2 L_2 \cos(\theta_2)\right]d}{2I}\right)^2 + 3\left(\dfrac{Td}{2I}\right)^2\right\}^{\frac{1}{2}} \quad (25)$$

**Table 6** summarizes the 11 input variables:

Table 6. Explanations for the 11 input variables

| NO. | Input variables | Form of distribution | Distribution parameter 1(mean) | Distribution parameter 2(std[1]) | Unit |
|---|---|---|---|---|---|
| $X_1$ | $t$ | Normal | 5 | 0.1 | mm |
| $X_2$ | $d$ | Normal | 42 | 0.5 | mm |
| $X_3$ | $L_1$ | Uniform | 119.75(lb[2]) | 120.25(ub[3]) | mm |
| $X_4$ | $L_2$ | Uniform | 59.75(lb) | 60.25(ub) | mm |
| $X_5$ | $F_1$ | Normal | 3 | 0.300 | kN |
| $X_6$ | $F_2$ | Normal | 3 | 0.300 | kN |
| $X_7$ | $P$ | Gumbel | 12 | 1.2 | kN |
| $X_8$ | $T$ | Normal | 90 | 9 | N-m |
| $X_9$ | $\theta_1$ | Uniform | $-\pi/3$ | $\pi/3$ | - |
| $X_{10}$ | $\theta_2$ | Uniform | $-4\pi/5$ | $2\pi/5$ | - |
| $X_{11}$ | $S_y$ | Normal | 220 | 22 | MPa |

Note: std[1]—standard deviation; lb[2]—Lower bound for uniform distribution; ub[3]—upper bound for the uniform distribution.

The simulation was carried out with two horizontal lines with 101 and 1001 samples respectively, and the obtained modeling RAAE is shown in **Table 7,** and the corresponding error graphs are shown in **Figures 5** and **6**.

75

International Journal of Computer Science & Information Technology (IJCSIT) Vol 15, No 3, June 2023

Table 7. Results of Cut-HDMRs

| HDMRs | RAAE | Samples | Time |
| --- | --- | --- | --- |
| PCE-HDMR | 1.022 | 101 | **0.079233** |
| Kriging-HDMR | 1.069 | 101 | 0.085015 |
| **PC-Kriging-HDMR** | **0.801** | **101** | 0.079864 |
| PCE-HDMR | 0.230 | 1001 | **9.475156** |
| Kriging-HDMR | 0.154 | 1001 | 9.833141 |
| **PC-Kriging-HDMR** | **0.064** | **1001** | 9.833141 |

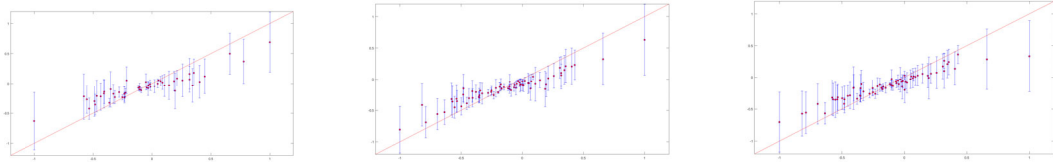

(a)PCE-HDMR(RAAE0.1022) (b)Kriging-HDMR(RAAE1.069) (c)PC-KrigingHDMR(RAAE0.801)

**Figure 5.** Results of three Cut-HDMRs when the number of samples is 101

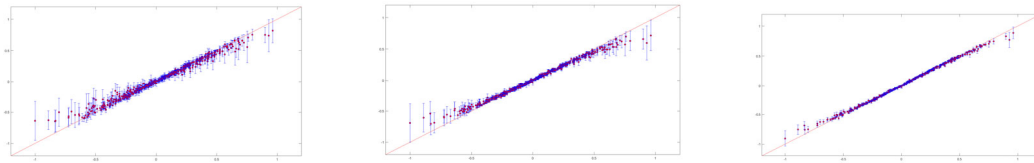

(a)PCE-HDMR(RAAE0.230) (b)Kriging-HDMR(RAAE0.154) (c) PC-Kriging-HDMR(RAAE0.064)

**Figure 6.** Results of three Cut-HDMRs when the number of samples is 1001

The results show the performance of the proposed PC-Kriging-HDMR combined with the adaptive sequential sampling strategy is slightly better than that of PCE-HDMR and Kriging-HDMR under the same sampling number cost.

In order to fully reflect the application advantages of PC-Kriging-HDMR, the sensitivity analysis will be carried out with reference to [24]. The specific results are shown in **Tables 8,9** and **Figure 7**.

Table 8. Sensitivity analysis results of single input variable

| Sensitivity ranking | Input variables(11) | Index of sensitivity(S) |
| --- | --- | --- |
| 1 | $X_8(T)$ | 0.6695 |
| 2 | $X_2(d)$ | 0.0111 |
| 3 | $X_{11}(S_y)$ | 0.0021 |
| 4 | $X_7(P)$ | 3.0141e-04 |
| 5 | $X_6(F_2)$ | 6.2951e-05 |
| 6 | $X_3(L_1)$ | 5.7746e-05 |
| 7 | $X_4(L_2)$ | 5.7746e-05 |
| 8 | $X_9(\theta_1)$ | 5.7746e-05 |
| 9 | $X_{10}(\theta_2)$ | 5.7746e-05 |
| 10 | $X_5(F_1)$ | 4.1750e-05 |
| 11 | $X_1(t)$ | 1.1506e-05 |

Table 9. Sensitivity analysis results of the coupling input variables

| Sensitivity ranking | Input variables | Index of sensitivity(S) |
| --- | --- | --- |
| 1 | $X_6(F_2)$, $X_8(T)$ | 2.8653e-04 |
| 2 | $X_8(T)$, $X_{11}(S_y)$ | 2.7792e-04 |
| 3 | $X_2(d)$, $X_8(T)$ | 2.7687e-04 |
| 4 | $X_3(L_1)$, $X_8(T)$ | 2.5937e-04 |
| 5 | $X_4(L_2)$, $X_8(T)$ | 2.5937e-04 |
| 6 | $X_8(T)$, $X_9(\theta_1)$ | 2.5937e-04 |
| 7 | $X_8(T)$, $X_{10}(\theta_2)$ | 2.5937e-04 |
| 8 | $X_5(F_1)$, $X_8(T)$ | 2.2353e-04 |
| 9 | $X_1(t)$, $X_8(T)$ | 1.2415e-04 |
| …… | …… | …… |
| 55 | $X_2(d)$, $X_6(F_2)$ | 4.7585e-07 |

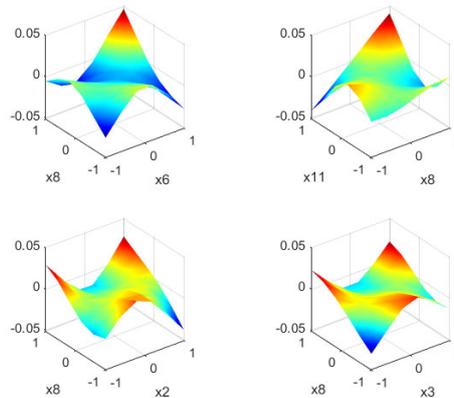

**Figure 7.** Variable coupling diagram (partial display)

From the perspective of a single variable, Table 8 shows that the torque $T$ has the greatest influence on the displacement response and stress distribution, while $\theta_2$ has the least influence. From the perspective of coupled variables, Table 9 and Figure 7 show that the coupling effect of arm tube width $t$ and torsion $T$ has the greatest impact on the response, thus quantifying that PC-Kriging-HDMR can directly reveal the corresponding main effects and interaction effects between variables, which indicates it has certain practical significance for sensitivity analysis.

## 6. CONCLUSIONS

By decoupling the variables of the high-dimensional multi-parameter problem, the black-box becomes transparent, aiming to establish a high-precision surrogate model on the basis of reducing the sampling cost. In this paper, a PC-Kriging surrogate model is embedded into the high-dimensional model representation framework Cut-HDMR, and a multi-stage adaptive sequential sampling strategy is designed. Through a series of numerical tests and a practical application of cantilever reliability modeling, the results show that: (1) The number of samples for PC-Kriging-HDMR modeling increases polynomially rather than exponentially with the expansion of the parameter space, which greatly reduces the computational cost; (2) Cut-HDMRs based on the single-surrogate model is not significantly superior to other methods, while PC-Kriging-HDMR performs better than Kriging-HDMR and PCE-HMDR in modeling accuracy and efficiency, which shows strong robustness; (3) Through the multi-stage adaptive sequential sampling strategy based on adaptive proportion criterion and central-baesd maximum entropy criterion, the layout of test sites is optimized and the convergence speed of the component surrogate model is accelerated.

This paper has made a exploration of the high-dimensional multi-parameter decoupling problem, and there are still the following directions to be further explored in the future:(1) The PC-Kriging-HDMR approximate modeling technique proposed in this paper mainly focuses on the truncation of low-order component terms, ignoring the errors that may be caused by high-order coupled terms. To further improve the accuracy of the model, it is necessary to correctly estimate the above errors and add appropriate correction terms; (2) The design of sequential sampling strategy needs to be improved. Efficient adaptive sequential sampling methods usually need to consider three parts: local search, global exploration, and the tradeoff between local and global. How to self-adaptive sampling according to the model output information still deserves further study; (3) Under the framework of PC-Kriging-HDMR, the idea of multi-fidelity modeling can also help to explore the range of system behavior and parameter variation, and improve the efficiency of analysis and optimization.

## ACKNOWLEDGEMENTS

This work is supported by National Natural Science Foundation of China (No.12201656)